\pdfoutput=1

\documentclass[11pt]{article}

\usepackage[]{EMNLP2023}

\usepackage{times}
\usepackage{latexsym}

\usepackage[T1]{fontenc}

\usepackage[utf8]{inputenc}

\usepackage{microtype}

\usepackage{inconsolata}

\title{ Current state of LLM Risks and AI Guardrails }

\author{
  \textbf{Suriya Ganesh Ayyamperumal},\\
  Carnegie Mellon University, sayyampe@andrew.cmu.edu \\
  \textbf{Limin Ge},\\
  Carnegie Mellon University, liminge@andrew.cmu.edu \\
  }

\begin{document}
\maketitle
\begin{abstract}

Large language models (LLMs) have become increasingly sophisticated, leading to widespread deployment in sensitive applications where safety and reliability are paramount.  However, LLMs have inherent risks accompanying them, including bias, potential for unsafe actions, dataset poisoning, lack of explainability, hallucinations, and non-reproducibility. These risks necessitate the development of "guardrails" to align  LLMs with desired behaviors and mitigate potential harm.

This work explores the risks associated with deploying LLMs and evaluates current approaches to implementing guardrails and model alignment techniques. We examine intrinsic and extrinsic bias evaluation methods and discuss the importance of fairness metrics for responsible AI development.  The safety and reliability of agentic LLMs (those capable of real-world actions) are explored, emphasizing the need for testability, fail-safes, and situational awareness.

Technical strategies for securing LLMs are presented, including a layered protection model operating at external, secondary, and internal levels. System prompts, Retrieval-Augmented Generation (RAG) architectures, and techniques to minimize bias and protect privacy are highlighted.

Effective guardrail design requires a deep understanding of the LLM's intended use case, relevant regulations, and ethical considerations. Striking a balance between competing requirements, such as accuracy and privacy, remains an ongoing challenge. This work underscores the importance of continuous research and development to ensure the safe and responsible use of LLMs in real-world applications.

\end{abstract}

\section{Introduction}

There has been a huge surge in deployments and utilization of LLMs, with Trillions of parameters, with a Mixture of Experts (MoE) based architecture \cite{gormley2019mixture} . The exact number of parameters for the frontrunning proprietary models such as ChatpGPT-4 from OpenAI and Gemini1.5-Ultra from Google have been guarded secretly and haven't been revealed publicly. Alongside proprietary LLMs with Large parameters, there has also been a surge in open source and proprietary LLMs that can be deployed in a single computer with as low as 3 Billion parameters to 100s of Billion parameters.

Since this is a very fast evolving field, multiple online sources have been cited in this paper.

LLMs are probabilistic next word prediction models, which are being increasingly deployed in a wide range of critical tasks such as law cite[], medical health record management cite[], etc. These models are still prone to the above stated risks and exposes the user as well as the maintainer of these widely deployed models to a wide range of Risks and challenges including but not limited to Bias and Fairness, Dataset Poisoning, Explainability, Hallucinations and Privacy. 

Since LLMs contain trillions of parameters, it is hard to predict or prove how or what a model would provide as an output in a given moment. These provide significant problems over how a model is safely deployed. There have been multiple examples of real-world deployments of LLMs having unexpected runaway behaviours, resulting in monetary as well as reputational loss.

One way of reducing these risks is to implement safety protocols intended to control the behaviour of these LLMs. These are provided in the form of Programmable "guardrails" by model developers – algorithms that monitor the inputs and outputs of an LLM.  Guardrails can, for instance, stop LLMs from processing harmful requests or modify results to be less dangerous or conform to the deployers specific requirement on morality.

Effective guardrails design is difficult and nuanced. Most Difficulty in Designing a good guardrails often is defining the requirements and expectation from the ML model. For example, Regulations vary between fields, country and region. Ethical requirements like fairness or avoiding offensive responses are hard to define concretely in an actionable way.

Despite all this, when deploying an LLM alongside or integrated with a conrete application. Even common sense requirements such as reducing hallucination \cite{xu2024hallucination} , toxicity \cite{wen2023unveiling} and biases\cite{taubenfeld2024systematic} are non-trivial tasks that are being explored. The requirement to cater to a specific utilitarian use case compounds the problem, making it harder. Often times, Model developers have competing requirements. For example, model developers need to select between Accuracy vs Privacy of the model, if the training data is scrubbed of private data and privacy friendly Machine Learning algorithms are used, accuracy takes a hit, resulting in worse performing models.

An Optimal guardrail would conform the model to the required task and prevent damage when the model strays out of its path. There is no kill all type solution to these problems. Currently, deployers of LLMs are taking a multi-pronged path where each special condition is tagged and handled programmatically. There are also attempts to build expert models optimized for specific outputs.

The key objective of this work was to:
\begin{itemize}
    \itemsep0em
    \item Enumerate the Risk exposure when working with and deploying Large Language Models.
    \item Evaluate the current state of Technical and Implementation Challenges of LLM Guardrails and Model Alignment works to provide Safety Guarantees during deployment.
\end{itemize}

\section{ Large Language Model Risks}

\subsection{Bias}

Bias in LLMs \cite{yeh2023evaluating} \cite{von2023assessing} refer to systematic errors or deviations in prompt outputs. that favor certain individuals or groups over others, often based on sensitive characteristics like race, gender, or socioeconomic status. This generally due to biased training data, which reflects existing societal prejudices, which was present in the training material that these LLMs were trained on.
It has been shown that LLMs can also be intentionally biased by prompts, posing a security/deployment risk for LLMs. \cite{badyal2023intentional}

\subsubsection{Intrinsic Bias Evaluation}
Intrinsic Biases are evaluated by looking at the internal word embedding cite[] representation of an LLM without directly looking at the outputs.
These include Word Embedding Association Test (WEAT) \cite{li2021self} target words (e.g., gender, race) and attribute words (e.g., "pleasant", "unpleasant"). High association scores may signal bias.
SECT (Sentential Embedding Association Test) \cite{prior2008word} is another measure that is Similar to WEAT but uses sentences instead of single words for more nuanced analysis.

\subsubsection{Extrinsic Bias Evaluation}

These evaluate bias by utilizing task specific benchmarks, depending on the context of the LLMs deployment.
Some popular datasets to detect biases are,

Winogender \cite{rudinger-EtAl:2018:N18}: A dataset assessing gender bias in pronoun resolution tasks.

StereoSet \cite{nadeem2020stereoset}: Measures bias across professions, religion, etc., in various NLP tasks like sentiment analysis and question answering.

Counterfactual Evaluation: Involves generating text with altered protected attributes (e.g., changing gender) and observing how the model's output changes.

\subsection{Fairness}

Fairness in machine learning \cite{pessach2022review} seeks to ensure that models make decisions without unjust discrimination. Achieving fairness requires careful attention to how data is collected, how models are trained, and how their outcomes are interpreted. It aims to create algorithms that treat all individuals impartially and promote equitable outcomes.

\subsection{Agentic Systems}

Agentic Large Language Models are systems with the ability to take real world actions from the digital environments. These are models that are able to move beyond simple text generation and include searching for information online , controlling smart devices cite[], create data analysis strategies \cite{yang2024matplotagent}, etc.

\subsection{Safety and Reliabilty}

Safety and Reliability is Pivotal when deploying LLMs that are able to have real world impact. Treating LLMs as a probabilistic black box would have real world impact, especially when these agents start making decisions that cross the boundary from digital to physical. \cite{chan2023harms}

\subsubsection{Testability}

This involves subjecting the LLMs to a wide range of simulated situations, including those that are unusual or extreme (edge cases). The goal is to identify any potential responses from the LLM that could trigger unsafe actions by the system and address those issues before real-world deployment.
A good implementation of Guardrails would be able to handle this.

\subsubsection{Fail Safes}
These are essential backup systems designed to take control if the LLM malfunctions.  Fail-safes might include manual human overrides or programmed instructions to automatically shut down certain processes to prevent the system from causing harm or damage.

\subsubsection{Situational Awareness}

For an LLM to reliably guide a system's actions, it needs continuous updates about the state of its surroundings. But, by design LLMs are stateless next word prediction machines, to work around this problem, the state is provided in the Prompt Context window cite[] This would include things gathered by sensors (cameras, location data, temperature monitors, etc.) to allow the LLM to make informed and contextually appropriate decisions.

\subsection{ Poisoned Datasets }

Large language models (LLMs) are trained on massive amounts of text data.  Poisoned datasets occur when this training data intentionally includes malicious, misleading, or harmful content. The goal of poisoning is to manipulate the model's output, causing it to generate biased, offensive, or unsafe text.

Poisoned datasets present significant risks. A compromised model might perpetuate harmful stereotypes, spread misinformation, or be used to generate material  that targets or  degrades specific groups.  The consequences can range from reputational damage for the organization deploying the LLM to more direct threats to users if the model is used for sensitive tasks. As shown by someonetlal cite[], LLMs can be jailbroken with specific wakewords by poisoning the dataset that they were trained on.

\subsection{ Explainability and Interpretability}

LLMs, due to their complex neural network architectures, are often referred to as black boxes. These  LLMs contain Trillions of parameters making it extremely difficult to understand exactly how they arrive at their decisions or why they generate a particular piece of text. The lack of explainability and interpretability makes it hard to trust an LLM's output, especially in high-stakes situations.

When deployed in situations where there need to be clear justifications, such as medical cite[], and legal cite[] an unexplainable model might be unusable, regardless of its accuracy. Additionally, the lack of interpretability poses a challenge in debuggin and improving the mdoel. The State of the art GPT-4, became "lazy" during the december 2024, and even ML scientists working on the forefront of LLMs couldn't figure out the issue cite[]

\subsection{ Hallucinations } \label{hallucinations}

Hallucinations are a major concern when it comes to large language models (LLMs).  Fabricated output can be surprisingly convincing, often containing plausible  sounding details or adhering to grammatical rules. This has lead to real world impacts such as lawyers getting caught citing non present legal precedence cite[]

The problem arises because LLMs are trained on massive datasets of text and code, and they learn to identify patterns and statistical relationships within that data. However, they don't inherently understand the real world or possess the ability to discern truth from fiction. This can lead the model to create  seemingly factual responses that are entirely  made up.

Hallucinations pose a significant risk because they erode trust in LLMs. If a user  can't be certain  the information they receive is accurate, the model becomes unreliable. This is especially dangerous in domains where factual accuracy is crucial, such as healthcare or finance.  Furthermore,  hallucinations can be  malicious,  as they could be used to  spread misinformation or create fake content that sways public opinion.  Therefore, mitigating hallucinations is an ongoing area of research in the field of large language models.

\subsection{ Non reproducability }

Non-reproducability in LLMs is their tendency to generate different outputs for the same input, irrespective of the timeframe. This inconsistency stems from the stochastic nature of LLMs cite[]. Stochastic elements, essentially random factors, are woven into an LLM's training process, and influence its internal calculations. 

This lack of consistency poses significant challenges when non-reproducability undermines real-world applications. Evaluating LLM performance becomes unreliable when benchmark results fluctuate. More importantly. In scenarios where consistent responses are crucial, like chatbots used in customer service, unpredictable outputs can lead to user frustration and a breakdown in trust. To ensure the dependability of LLMs, researchers are actively exploring methods to mitigate non-reproducability and ensure a more consistent user experience.

\subsection{ Privacy and copyright }

LLMs are known to memorize complete texts and reproduce private data present in the training data. This could be data that was proprietary or inadvertently part of a data leak. There have been adversarial way to fetch

\section{ Strategies in Securing Large Language models }

Securing large language models (LLMs) is a multifaceted endeavor due to their unique vulnerabilities and the sensitive data they often process.  Protecting these models requires a combination of robust access controls to prevent unauthorized use, careful monitoring of inputs and outputs to detect malicious prompts or harmful responses. It's essential to stay vigilant about potential biases embedded within the training data and actively work to mitigate their effects.  Additionally, it is also important to provide a way to validate the models output to be valid or not.

\begin{table*}\label{table:risk-mitigation-table}
\centering
\begin{tabular}{|l|l|l|l|}
\hline
& \textbf{Gatekeeper Layer}                                                                                                                                                                                                                                                                                 & \textbf{\begin{tabular}[c]{@{}l@{}}Knowledge Anchor \\ Layer\end{tabular}}                                                                                & \textbf{Parametric Layer}                                                                                                                                                                                 \\ \hline
\textbf{\begin{tabular}[c]{@{}l@{}}Response \\ Reliability\end{tabular}}      & \begin{tabular}[c]{@{}l@{}}- In Context Learning \\ \cite{10.1145/3571730}\\ \cite{min2022rethinking}\\ \\ - Instruction Tuning\\ \cite{peng2023instruction}\\ \\ \\ - fewshot prompting\\ \cite{liu2024stablept}\end{tabular}                                                                            & \begin{tabular}[c]{@{}l@{}}- RAG with short \\ Cosine Distance\\ \cite{asai2024reliable}\\ \\ \\ - Corrective RAG\\ \cite{yan2024corrective}\end{tabular} & \begin{tabular}[c]{@{}l@{}}- Task/domain specific \\ finetuning\\ \cite{fernandez2022fine}\\ \\ \\ - Fine tuned specialist \\ models\\ \cite{ngiam2018domain}\\ \cite{schimanski2024towards}\end{tabular} \\ \hline
\textbf{Hallucination}                                                        & \begin{tabular}[c]{@{}l@{}}- Self consistency \\ check\\ \cite{lecky1945self}\\ \cite{min2023beyond}\\ \cite{chen2023universal}\end{tabular}                                                                                                                                                              & \begin{tabular}[c]{@{}l@{}}- Knowledge Graphs\\ \cite{abu2024knowledge}\\ \cite{vamsi2024human}\end{tabular}                                              & \begin{tabular}[c]{@{}l@{}}- Unfamiliar finetuning\\ \cite{kang2024unfamiliar}\end{tabular}                                                                                                               \\ \hline
\textbf{Bias}                                                                 & \begin{tabular}[c]{@{}l@{}}- In context examples \\ to counter balance the bias\\ \cite{dwivedi2023breaking}\\ \cite{schubert2024context}\\ \\ - Fairness guided \\ few-shot prompting\\ \cite{NEURIPS2023_8678da90}\\ \\ \\ - Validate the response \\ for bias\\ \cite{huang2024empirical}\end{tabular} & \begin{tabular}[c]{@{}l@{}}- RAG with \\ couterbalanced data\\ \cite{shrestha2024fairrag}\end{tabular}                                                    & \begin{tabular}[c]{@{}l@{}}- Fine tune with synthetic \\ data to balance\\ \cite{van2021decaf}\end{tabular}                                                                                               \\ \hline
\textbf{\begin{tabular}[c]{@{}l@{}}Protection from \\ Adversary\end{tabular}} & \begin{tabular}[c]{@{}l@{}}- Prompt Injection\\ \cite{choi2022prompt}\\ \\ - Membership Inference \\ attack\\ \cite{shokri2017membership}\\ \\ - Hardened System \\ Prompts.\\ \cite{zheng2023helpful}\end{tabular}                                                                                 & \begin{tabular}[c]{@{}l@{}}- Tighter \\ cosine similarity\\ \cite{rahutomo2012semantic}\end{tabular}                                                      & \begin{tabular}[c]{@{}l@{}}- Adversarial finetuning\\ \cite{jeddi2020simple}\\ \cite{chen2020adversarial}\end{tabular}                                                                                    \\ \hline
\textbf{Unknowability}                                                        & \begin{tabular}[c]{@{}l@{}}- Validate response with \\ external truth.\\ \cite{zhuang2024toolqa}\\ \\ - Master model with \\ prompt unknowability\\ \cite{zheng2024judging}\end{tabular}                                                                                                                  & \multicolumn{1}{c|}{-}                                                                                                                                    & \begin{tabular}[c]{@{}l@{}}- Counterexample finetuning\\ \cite{jha2023counterexample}\end{tabular}                                                                                                        \\ \hline
\end{tabular}
\caption{
Risks and mitigation strategies at different layers of an LLM
}
\end{table*}

\subsection{ Layered Protection model }

Broadly, guardrails are created at different layers of model access. The least privileged access method to an LLM is through Network APIs as provided by OpenAI, Anthropic and other model hosters.
In these cases, the attacker/users don't get access to the model architecture, Probability distribution of tokens, or training data. In these cases, securing and guardrailing can be done at each of these different layers. As described in \ref{table:risk-mitigation-table} each risk has multiple mitigation strategies at different layers.

\subsubsection{ GateKeeper Layer }

System prompts are instructions or guidelines given to a LLM to shape its behavior and responses. They act like a compass, steering the LLM's focus and defining parameters for its output. System prompts can establish the LLM's role (e.g., helpful assistant, creative writer), set the desired tone (formal, playful), provide background context, specify output formats, or even incorporate rules to keep the LLM's responses safe and appropriate. By thoughtfully designing system prompts, developers can significantly influence how the LLM behaves, tailoring its responses to specific applications and minimizing the risk of undesirable output.

To mitigate risks associated with malicious prompts and providing guarantees about the robustness of the LLM. A key protection layer lies outside the LLM itself, serving to detect potential attacks such as prompt injection. These attacks aim to manipulate the model's behavior, generating  biased or unintended outputs. By identifying malicious prompts during the tokenization stage, this layer can prevent harmful consequences like data breaches.

To bolster these defenses, the Gatekeeper protection layer may employ strategies like on-the-fly prompt rephrasing and rewriting. This preserves the core intent of the prompt while neutralizing potential areas of vulnerability.  This layer leverages specialized models and agents to comprehensively evaluate incoming prompts. These evaluation agents, while often smaller than the primary LLM, offer targeted expertise for prompt analysis.

It's important to acknowledge that even the most sophisticated protection measures can occasionally be circumvented.  Continuous research and development are essential to stay ahead of evolving attack strategies and to ensure that LLMs remain secure and aligned with their intended purpose.

\subsubsection{ Knowledge Anchor Layer }

This is the layer where the decision on which tokens should be included in the model is decided. This stage is primarily used to protect from risks such as Hallucinations and Ground the model to reality. This is done by provided by creating Vector data

LLMs are known for their tendency to "hallucinate" or generate responses that are plausible but factually inaccurate.  RAG architectures address this vulnerability by grounding LLM responses in provided source material. Instead of simply inventing information, the LLM works with a curated knowledge base, increasing the likelihood of accurate and reliable output. This makes it more difficult for attackers to exploit LLM weaknesses and spread misinformation.

One risk associated with LLMs is that they can sometimes leak private information that was present in their training data. RAGs mitigate this by reducing the LLM's direct reliance on internal memorized training data. When an LLM is instructed to retrieve and process information from external documents, it lessens the chance of accidentally regurgitating sensitive information that may have been part of its original training. This helps protect privacy and lessens the potential for data breaches.

\subsubsection{ Parametric Layer }

These are the changes done to the model parameters either by finetuning, or addressing problems during the pretraining process of the model.

Researchers are actively exploring techniques to minimize biases present in large language models (LLMs). Methods range from fine-tuning existing models on carefully curated data to modifying the pre-training process itself.  These approaches aim to address biases related to stereotypes, cultural insensitivity, and unfair representations of various groups.

Protecting the privacy of data used to train LLMs and the privacy of model outputs is crucial. Differential Privacy (DP) is a widely used framework used for this purpose. Techniques include modifying the fine-tuning process with DP principles, applying Local Differential Privacy (LDP), and exploring the concept of contextual privacy to tailor protections based on how sensitive information is within a specific context.

Enhancing the safety of LLMs can be done by adding adversarial examples to training data to make the models more robust to harmful inputs. Researchers focus on refining the reinforcement learning with human feedback (RLHF) process, using mechanisms to filter out harmful responses and modifying loss functions.  However, directly applying traditional robustness techniques can be challenging for LLMs due to complexities like catastrophic forgetting.

\section{ Challenges in Implementing Guardrails }

\subsection{ Flexibility vs Stability }

AI systems enable the flexibility to adapt, learn, and evolve in response to new data and changing circumstances. This is crucial for their continued effectiveness in dynamic real-world environments. But, stability is essential to ensure that the AI system operates within predictable boundaries, avoiding unintended consequences and maintaining alignment with ethical principles. Overly rigid guardrails limits the potential benefits of AI, while insufficient safeguards expose systems to risks such as bias, safety hazards, and misuse. Finding an optimal equilibrium between these in both language tasks as well as Agentic tasks, is an ongoing conversation between industry leaders and others.

\subsection{ Emergent Complexity }

When building with non-deterministic systems, there are multiple fallbacks and scaffoldings built to provide a certain level of confidence over the system. But, these are strung together with texts and prompts that are nebulous and hard to evaluate both qualitatively and quantitatively.
Creating a deterministic software system when the underlying systems are non-deterministic is an ongoing challenge for deployers of these LLM systems.

\subsection{ Unclear goals and metrics }

Often times when building LLM systems, the requirements on what is expected of the system is unclear. For Example, cosine similarity is by far the most popular metric utilized by all LLM application to validate their response. But, cosine similarity is shown to be very foggy and unable to predict the next steps.

\subsection{ System Testability and Evolvability }

When building with LLMs it is hard to prove that the system will function as designed and not go beyond the conditional assumptions of the developer. Often it is the case that the developer does not even control the model which is doing the information processing.
It is also hard to evaluate the model in terms of performance increase or degradation over time.

\subsection{ Cost }
A common technique in detecting and preventing malicious input or biased response, is to use a "Judge" LLM to evaluate and rate the responses. This is prohibitively costly both financially as well as in time to response.
It has been shown that smaller specialized "Judge" models perform much better than large language models with trillions of parameters in evaluating LLM responses.

\section{ Open Source Tools }
To enable building more stable LLM based application, there has been a surge in open source tools specifically built to enable guardrailing. These tools all take a slightly different approach to solving these problems.

\subsection{ Nemo-Guardrails }

Nemo-Guardrails \cite{rebedea-etal-2023-nemo} has a Domain Specific Language (DSL) called Colang  . It is possible define new and reuse predefined guardrails in the library. The decision on whether an evaluation needs to be done or not is decided based on the colang configuration. It is possible to setup separate processes at every step of an Input/Output pipeline. This also includes manipulating output from a Retrieval-Augmented-Generation System as well.

Nemo-Guardrails heavily utilizes LLMs such as Chat-gpt 4 and to evaluate and decide what next steps to perform. This could pose a potential risk, where it has recently been shown that models are biased towards their own outputs. Interestingly Nemo lacks controls on the cost of running a request, hence there could be cases where a simple inference could cost a lot more than expected.

\subsection{ LLamaGuard }

LLama Guard \cite{inan2023llama} is a solution created by Meta based on llama2-7b model for classifying Input, Output pairs into categories such as sensitivity. Llama Guard performed better than other traditional methods of classification because of its ability to understand semanticity of words and sentences.

LlamaGuard was built by  creating a dataset of ideal input/ouput pairs for certain case and then fine-tuning a foundational LLama-7B. The annotation was done by Meta's red team.

\subsection{Guardrails AI}

Guardrails AI is similar to Nemo-Guardrails in that, it also defines a Domain Specific Language called RAILS which is as a customized XML format, it is also able to do fuzzy matching for output semantic correctness. Guardrails AI is also able to leverage other LLMs to evaluate and make decisions on the outputs and inputs.
XML has been shown to be better understood, processed, and followed by LLMs because of how tokenization works.

\section{ Limitations }

LLM utilization and deployment is a fast evolving field, with a lot of research and development being done concurrently. While, our exploration has been depth first with a fair amount of breadth. Some of the strategies might get obsolete or proven wrong, very fast.

\section{ Conclusion }

LLMs offer exceptional potential for transformation across various industries. However, their inherent biases, safety concerns, and potential for misuse necessitate the development of robust guardrails. This work has explored the risks associated with deploying LLMs and the current state of technical strategies and model alignment techniques.

We examined methods for bias evaluation and the significance of fairness metrics. For agentic LLMs, we highlighted the need for rigorous testing, fail-safes, and situational awareness to ensure safe and reliable operation in real-world environments.  Technical approaches to securing LLMs, including layered protection models, system prompts, RAG architectures, and bias mitigation, were discussed.

Crucial challenges remain in implementing these guardrails. Finding the optimal balance between flexibility and stability in AI systems is essential to preserve both their adaptability and ethical alignment.  Clear definitions of goals and metrics, testability, and cost optimization are further areas requiring focus. Open-source tools offer promising avenues for building more reliable LLM-based applications.

Effective guardrail design demands a deep understanding of an LLM's intended use, relevant regulations, and ethical considerations. Despite the complexities, developing reliable safeguards is paramount for maximizing LLMs' benefits and minimizing their potential harms. Continued research, development, and open collaboration across the field are vital to ensuring the safe, responsible, and equitable use of LLMs in the years to come.

\bibliographystyle{acl_natbib}
\bibliography{anthology}
\end{document}